\documentstyle[aps,epsf]{revtex}

\newcommand{\bq}{\begin{equation}}
\newcommand{\eq}{\end{equation}}
\newcommand{\bqn}{\begin{eqnarray}}
\newcommand{\eqn}{\end{eqnarray}}
\newcommand{\nb}{\nonumber}
\newcommand{\lb}{\label}

\begin{document}
 
\title{Gravitational Collapse of Cylindrical Shells
Made of Counter-Rotating Dust Particles} 
\author{Paulo R.C.T. Pereira\thanks{Email: terra@dft.if.uerj.br}}
\address{ Departamento de
Astrof\'{\i}sica, Observat\'orio Nacional~--~CNPq, 
Rua General Jos\'e Cristino 77, S\~ao Crist\'ov\~ao, 20921-400 
Rio de Janeiro~--~RJ, Brazil}
\author{Anzhong Wang\thanks{Email:
wang@dft.if.uerj.br} }  
\address{ Departamento de F\' {\i}sica Te\' orica,
Universidade do Estado do Rio de Janeiro,
Rua S\~ ao Francisco Xavier 524, Maracan\~a,
20550-013 Rio de Janeiro~--~RJ, Brazil}

\date{ May 31, 2000}

\maketitle

\begin{abstract}

The general formulas of a non-rotating dynamic thin shell that
connects two arbitrary cylindrical regions are given using Israel's
method. As an application of them, the dynamics of a
thin shell made of counter-rotating dust particles, which emits both
gravitational waves and massless particles  when it is expanding or
collapsing,  is studied. It is found that when the models represent a
collapsing shell, in some cases the angular momentum of the dust particles 
is strong enough to halt the collapse,  so that  a spacetime singularity is
prevented from forming, while in other cases it is not, and  
a line-like spacetime singularity is finally formed on the symmetry axis.

\end{abstract}
 
\vspace{0.82cm}


\section{Introduction}
 
 Gravitational collapse of a realistic body has been one of the most
thorny and important problems in Einstein's theory of
General Relativity. Due to the 
complexity of the Einstein field equations, the problem even in 
simple cases, such as, spacetimes with spherical symmetry, is still 
not well understood \cite{Josh1994}, and new phenomena keep
emerging \cite{Ch1993}. Particularly, 
in 1991 Shapiro and Teukolsky \cite{ST1991} studied  numerically the
problem of a dust spheroid, and found that only the spheroid is compact
enough, a black hole can be formed.  Otherwise, the collapse most
likely ends with a naked singularity. Later, Barrab\'es, Israel and Letelier
constructed an analytical model of a collapsing convex thin shell and found
that in certain cases no apparent horizons are formed, too \cite{BIL1991}.
Their results were soon generalized to  more general cases \cite{BGL1992}.
However, since in all the cases considered by them, the external gravitational
field of the collapsing shell is not known,  one cannot
exclude, similar to the ST case,  the formation of an outer event horizon
\cite{BIL1991,BGL1992}.  Since then, the gravitational collapse with
non-spherical symmetry has been attracting more and more attention. In
particular, by studying the collapse of a cylindrical shell that is made of
counter-rotating particles, Apostolatos and Thorne (AT)  showed analytically
that the centrifugal forces associated with an arbitrarily small amount of
rotation, by themselves, without the aid of any pressure, can halt the
collapse at some non-zero, minimum radius, and the shell will then oscillate
until it settles down at some final, finite radius, whereby a spacetime
singularity is prevented from forming on the symmetry axis \cite{AT1992}. Soon
after AT's work, Shapiro and Teukolsky studied numerically the gravitational
collapse of rotating spheroids, and found that the rotation indeed
significantly modifies the evolution when it is sufficiently large. However,
for small enough angular momentum, their simulations showed that spindle
singularities appeared to arise without apparent horizons, too. Hence, it is
possible that even spheroids with some angular momentum may still form naked
singularities \cite{ST1992}.

It should be noted that in the AT work it was considered only the case
where the shell has zero total angular momentum  and is momentarily
 static and radiation-free. In a realistic case, the spacetime has
neither cylindrical symmetry nor zero angular momentum, and gravitational and
particle radiations are always expected to occur. As a generalization of the
AT work, in this paper we shall consider the case where  cylindrical shell
radiates gravitational waves and massless particles, as it is collapsing,
while still keep the requirement that the total  angular momentum of the shell
be zero. Specifically, the paper is organized as follows: In Sec.
$II$, the formulas for a general dynamic timelike thin shell that connects two
arbitrary cylindrical regions are given, using Israel's formula
\cite{Israel1966}, while in Sec. $III$, a collapsing thin shell made of
counter-rotating dust particles is studied. To model the particle radiation
of the shell, we consider the case where the spacetime outside the shell is
described by an out-going radiation fluid  \cite{LW1994}.  The paper is ended
with Sec. $IV$, where our  main conclusions are presented.

\section{Dynamics of Cylindrical Thin Shells Without Rotation}

Both static \cite{stashell} and dynamic \cite{dyshell} cylindrical thin shells
with zero total angular momentum have been studied previously. However, in
most of these studies a specific form of metric was usually assumed, which is
valid only in some particular cases, such as, the spacetime is vacuum outside
and inside the shell \cite{AT1992}. In this section, we shall give a general
treatment that is valid for any dynamic timelike cylindrical thin shell,
connecting two arbitrary cylindrical regions.

To begin with,  let us consider the cylindrical spacetimes 
described by the metric,
\bq
\lb{eq1}
ds^{2}_{-} = f^{-}(t,r) dt^{2} - g^{-}(t,r) dr^{2} - h^{-}(t,r) dz^{2}
 - l^{-}(t,r) d\varphi^{2},
\eq 
where $\{x^{- \mu}\}\equiv \{t, r, z, \varphi\}, \; (\mu = 0, 1, 2, 3)$, are
the usual cylindrical coordinates. For the spacetimes to be 
cylindrical, several criteria have to be satisfied  \cite{PSW1996}. When the
symmetry axis is regular, those conditions are easily imposed. However, when
it is singular, it is still not clear  which kind of conditions should be
imposed \cite{MS1998}.

In general the spacetimes described by Eq.(\ref{eq1}) have two Killing
vectors, one is associated with the invariant translations along the symmetry
axis, $\xi_{(z)} = \partial {z}$, where $z$ is the Killing coordinate length
with $ - \infty < z < + \infty$, and the other is associated with the
invariant rotations  about the axis, $\xi_{(\varphi)} =  \partial {\varphi}$
with $0 \le \varphi \le 2\pi$, where the hypersurface $\varphi = 0$ is
identical with the one $\varphi = 2\pi$. Clearly, for the metric given above,
the two Killing vectors are orthogonal. Consequently, the  metric represents
spacetimes without rotation, and the polarization of gravitational waves  has
only one degree of freedom \cite{Thorne1965,MTW1973}.

Assume that a given spacetime is divided by a hypersurface $\Sigma$ into two
regions, say, $V^{\pm}$, where the region $V^{-}$ is described by the metric
(\ref{eq1}), while the region $V^{+}$ is described by the metric
\bq
\lb{eq1a}
ds^{2}_{+} = f^{+}(T,R) dT^{2} - g^{+}(T,R) dR^{2} - h^{+}(T,R) dz^{2}
 - l^{+}(T,R) d\varphi^{2},
\eq 
where $\{x^{+ \mu}\}\equiv \{T, R, z, \varphi\}, \; (\mu = 0, 1, 2, 3)$, is
another set of the cylindrical coordinates. The hypersurface $\Sigma$
in the coordinates $x^{\pm\mu}$ is given, respectively, by
\bq
\lb{eq3} 
r = r_{0}(t),\;\;\; R= R_{0}(T).  
\eq
On the surface, the metrics (\ref{eq1}) and (\ref{eq1a}) reduce, respectively,
to
\bqn
\lb{eq4}
\left. ds^{2}_{-}\right|_{r= r_{0}(t)} &=& \left[f^{-}\left(t,r_{0}(t)\right) 
-  g^{-}\left(t,r_{0}(t)\right) {r'}_{0}^{2}(t)\right] dt^{2} 
- h^{-}\left(t,r_{0}(t)\right) dz^{2}
 - l^{-}\left(t,r_{0}(t)\right) d\varphi^{2},\nb\\
\left. ds^{2}_{+}\right|_{R= R_{0}(T)} &=& \left[f^{+}\left(T,R_{0}(T)\right) 
-  g^{+}\left(T,R_{0}(T)\right) {R'}_{0}^{2}(T)\right] dT^{2} 
- h^{+}\left(T,R_{0}(T)\right) dz^{2}
 - l^{+}\left(T,R_{0}(T)\right) d\varphi^{2},
\eqn
where a prime denotes the ordinary differentiation with respect to the
indicated argument. In this paper, we shall consider only the case where
$\Sigma$ is timelike. Then, if we choose the intrinsic coordinates of the
hypersurface as $\{\xi^{a}\} = \left\{\tau, z, \varphi\right\},\; (a = 1, 2,
3)$, where $\tau$ denotes the proper time of the surface, we find that the
metric on the hypersurface can be written as, 
\bq
\lb{eq5}
\left. ds^{2}\right|_{\Sigma} = \gamma_{ab}d\xi^{a}d\xi^{b} = d\tau^{2} -
h(\tau)dz^{2} - l(\tau)d\varphi^{2}, 
\eq
where
\bqn
\lb{eq6}
d\tau  &=&
\left[f^{-}\left(t,r_{0}(t)\right)  -  g^{-}\left(t,r_{0}(t)\right)
{r'}_{0}^{2}(t)\right]^{1/2} dt  = \left[f^{+}\left(T,R_{0}(T)\right)  - 
g^{+}\left(T,R_{0}(T)\right) {R'}_{0}^{2}(T)\right]^{1/2} dT, \nb\\
h(\tau) &\equiv& h^{-}\left(t,r_{0}(t)\right) =
               h^{+}\left(T,R_{0}(T)\right),\;\;\;  
l(\tau) \equiv l^{-}\left(t,r_{0}(t)\right) =
l^{+}\left(T,R_{0}(T)\right),
\eqn
where the function dependence of $\tau$ on $t$ and $T$ is given by the first
equation. Note that in writing the above expressions, we had chosen $d\tau,\;
dT$ and $dt$, without loss of generality,  to have the same sign, and already
applied the first junction conditions,  
\bq
\lb{eq6a}
 \left.
ds^{2}_{-}\right|_{r= r_{0}(t)} = \left. ds^{2}_{+}\right|_{R= R_{0}(T)}.
\eq

It can be shown that the unit spacelike normal vector to the hypersurface
$\Sigma$ in the coordinates $x^{\pm\mu}$ is given, respectively, by 
\bqn
\lb{eq7}
n^{+}_{\mu}  &=& \left[\frac{f^{+}g^{+}}{f^{+} -
g^{+}{R'}^{2}_{0}(T)}\right]^{1/2} 
\left\{ - {R'}_{0}(T)\delta^{T}_{\mu} + \delta^{R}_{\mu}\right\},\nb\\
n^{-}_{\mu}  &=& \left[\frac{f^{-}g^{-}}{f^{-} -
g^{-}{r'}^{2}_{0}(t)}\right]^{1/2} 
\left\{ - {r'}_{0}(t)\delta^{t}_{\mu} + \delta^{r}_{\mu}\right\}.
\eqn
Then, the non-vanishing components of the extrinsic curvature tensor
$K^{\pm}_{ab}$, defined by\footnote{ Note that in this paper the definition
for the extrinsic curvature tensor is different from that of Israel by a ``$-$"
sign \cite{Israel1966}.}
\bq 
\lb{eq8}
K_{ab} = n_{\alpha}\left(
\frac{\partial^{2}x^{\alpha}}{\partial \xi^{a} 
\partial \xi^{b}}
+ \Gamma^{ \alpha}_{\beta\delta}\frac{\partial 
x^{\beta}}
{\partial \xi^{a}}\frac{\partial x^{\delta}}
{\partial \xi^{b}}\right),
\eq
are given by
\bqn
\lb{eq9}
K^{+}_{\tau\tau} &=& \frac{\left(f^{+}g^{+}\right)^{1/2}}
{2\left[f^{+} - g^{+}{R'}^{2}_{0}(T)\right]^{3/2}}
\left\{ - \frac{f^{+}_{,R}}{g^{+}} 
+ \left(\frac{f^{+}_{,T}}{f^{+}} 
 - 2 \frac{g^{+}_{,T}}{g^{+}}\right){R'}_{0}(T) \right.\nb\\ 
& & \left. + \left(2 \frac{f^{+}_{,R}}{f^{+}} -
\frac{g^{+}_{,R}}{g^{+}}\right){R'}^{2}_{0}(T) 
 + \frac{g^{+}_{,R}}{f^{+}}{R'}^{3}_{0}(T)
- 2{R''}_{0}(T)\right\},\nb\\
K^{+}_{zz} &=& \frac{1}{2}\left[\frac{f^{+}g^{+}}
{f^{+} - g^{+}{R'}^{2}_{0}(T)}\right]^{1/2}
\left\{ \frac{h^{+}_{,R}}{g^{+}} 
+ \frac{h^{+}_{,T}}{f^{+}} {R'}_{0}(T)\right\},\nb\\
K^{+}_{\varphi\varphi} &=& \frac{1}{2}\left[\frac{f^{+}g^{+}}
{f^{+} - g^{+}{R'}^{2}_{0}(T)}\right]^{1/2}
\left\{ \frac{l^{+}_{,R}}{g^{+}} 
+ \frac{l^{+}_{,T}}{f^{+}} {R'}_{0}(T)\right\},
\eqn
where $f^{+}_{,T} \equiv \partial f^{+}(T,R)/\partial T$ etc., and
$K^{-}_{ab}$ can be obtained from the above expressions by the replacement
\bq
\lb{eq10}
f^{+},\; g^{+},\; h^{+},\; l^{+},\; R_{0}(T),\; T,\; R \;\;\; \rightarrow
\;\;\; f^{-},\; g^{-},\; h^{-},\; l^{-},\; r_{0}(t),\; t,\; r. 
\eq

In terms of $K^{\pm}_{ab}$ and $ \gamma_{ab}$, the surface energy-momentum
tensor, $\tau_{ab}$, is defined as \cite{Israel1966},
\bq
\lb{eq11}
\tau_{ab} = \frac{1}{\kappa}\left\{ \left[K_{ab}\right]^{-} - \gamma_{ab}
[K]^{-}\right\},
\eq
where $\kappa[\equiv 8\pi G/c^{4}]$ is the Einstein constant, and
\bq
\lb{eq12}
\left[K_{ab}\right]^{-} \equiv K^{+} _{ab} - K^{-} _{ab},\;\;\;
[K]^{-} \equiv \gamma^{ab} \left[K_{ab}\right]^{-}.
\eq
Inserting Eq.(\ref{eq9}) and the corresponding expressions for $K^{-}_{ab}$
into Eq.(\ref{eq11}), we find that $\tau_{ab}$ can be written in the form
\bq
\lb{eq13}
\tau_{ab} = \rho w_{a}w_{b} + p_{z}z_{a}z_{b} 
+ p_{\varphi}\varphi_{a}\varphi_{b}, \; (a, b = \tau, \; z, \; \varphi),
\eq
where 
\bqn
\lb{eq14}
\rho &=& \frac{1}{\kappa}\left\{ 
\frac{\left[K_{zz}\right]^{-}}{h(\tau)} + 
\frac{\left[K_{\varphi\varphi}\right]^{-}}{l(\tau)}\right\},\nb\\
p_{z} &=& \frac{1}{\kappa}\left\{ 
\left[K_{\tau\tau}\right]^{-}  - 
\frac{\left[K_{\varphi\varphi}\right]^{-}}{l(\tau)}\right\},\nb\\
p_{\varphi} &=& \frac{1}{\kappa}\left\{ 
\left[K_{\tau\tau}\right]^{-}  - 
\frac{\left[K_{zz}\right]^{-}}{h(\tau)}\right\},
\eqn
and $w_{a},\; z_{a}$ and $\varphi_{a}$ are unit  vectors, defined as,
\bq
\lb{eq15}
w_{a} = \delta^{\tau}_{a},\;\;\;
z_{a} = h^{1/2}(\tau)\delta^{z}_{a},\;\;\;
\varphi_{a} = l^{1/2}(\tau)\delta^{\varphi}_{a}.
\eq
Clearly, the surface energy-momentum tensor given by Eq.(\ref{eq13}) can be
interpreted as representing a massive thin shell with its velocity
$w_{a}$, and principal pressures $p_{z}$ and $p_{\varphi}$, respectively, in
the direction, $z_{a}$ and $\varphi_{a}$, provided that it satisfies some
energy conditions \cite{HE1973}.

Using Eq.(\ref{eq5}) and Eqs.(\ref{eq13}) - (\ref{eq15}), one can show that
the conservation law on the hypersurface $\Sigma$ \cite{Israel1966},
\bq
\lb{eq16}
\tau^{b}_{a|b} = - \left[T^{+}_{\alpha\beta} n^{+\alpha}e^{+\beta}_{(a)}
- T^{-}_{\alpha\beta} n^{- \alpha}e^{- \beta}_{(a)}\right],
\eq
has only one non-vanishing component, which can be written as
\bq
\lb{eq17}
\frac{d\rho}{d\tau} + \frac{(\rho + p_{z})}{2h(\tau)}\frac{d h(\tau)}{d\tau}
+ \frac{(\rho + p_{\varphi})}{2l(\tau)}\frac{d l(\tau)}{d\tau} =
 - \left[T^{+}_{\alpha\beta} n^{+\alpha}e^{+\beta}_{(\tau)}
- T^{-}_{\alpha\beta} n^{- \alpha}e^{- \beta}_{(\tau)}\right],
\eq
where ``$|_{b}$" denotes the covariant differentiation with respect to the 
three-metric $\gamma_{ab}$, and $T^{\pm}_{\alpha\beta}$ are the energy-momentum
tensors calculated, respectively, in $V^{+}$ and $V^{-}$,  and 
 \bqn 
\lb{eq18}
e^{+\mu}_{(\tau)} &\equiv& \frac{\partial x^{+\mu}}{\partial \tau}
 = \left(f^{+} - g^{+}{R'}^{2}_{0}(T)\right)^{-1/2} 
\left\{\delta^{\mu}_{T} + {R'}^{2}_{0}(T)\delta^{\mu}_{R}\right\},\nb\\
e^{+\mu}_{(z)} &\equiv& \frac{\partial x^{+\mu}}{\partial z}
 =  \delta^{\mu}_{z},\;\;\;
e^{+\mu}_{(\varphi)} \equiv \frac{\partial x^{+\mu}}{\partial \varphi}
 =  \delta^{\mu}_{\varphi},\nb\\
e^{- \mu}_{(\tau)} &\equiv& \frac{\partial x^{-\mu}}{\partial \tau}
 = \left(f^{-} - g^{-}{r'}^{2}_{0}(t)\right)^{-1/2} 
\left\{\delta^{\mu}_{t} + {r'}^{2}_{0}(t)\delta^{\mu}_{r}\right\},\nb\\
e^{- \mu}_{(z)} &\equiv& \frac{\partial x^{- \mu}}{\partial z}
 =  \delta^{\mu}_{z},\;\;\;
e^{- \mu}_{(\varphi)} \equiv \frac{\partial x^{- \mu}}{\partial \varphi}
 =  \delta^{\mu}_{\varphi}.
\eqn

When no matter shell appears on the hypersurface $\Sigma$, we have $\tau_{ab}
= 0$, and the hypersurface represents a boundary surface \cite{Israel1966},
with the junction conditions being given by Eq.(\ref{eq6}) and
Eq.(\ref{eq17}).  The latter can be written in the form
 \bq
\lb{eq19}
\left. T^{+}_{\alpha\beta} n^{+\alpha}e^{+\beta}_{(\tau)}\right|_{\Sigma}
= \left. T^{-}_{\alpha\beta} n^{- \alpha}e^{-
\beta}_{(\tau)}\right|_{\Sigma},\; 
(\tau_{ab} = 0). 
\eq 

Once we have the general formulas, let us turn to consider their applications
to some specific cases.

\section{Gravitational Collapse of  Cylindrical Shells Made of Counter-Rotating
Dust particles} 
  
In this section, we shall consider the gravitational collapse of
a  cylindrical  shells made of counter-rotating dust particles. The shell
emits gravitational and particle radiations, when it is collapsing.  The metric
inside the shell will be chosen as that of Minkowski,
\bq
\lb{eq2.1}
ds^{2}_{-} = dt^{2} - dr^{2} - dz^{2} - r^{2}d\varphi^{2},
\eq
so that the symmetry axis is well defined and the local-flatness condition
is satisfied \cite{PSW1996}. The metric outside the shell will be
chosen as that representing out-going radiation fluid, given by \cite{LW1994} 
\bq 
\lb{eq2.2}
ds^{2}_{+} = e^{-b(\xi)}\left(dT^{2} - dR^{2}\right) - dz^{2} -
R^{2}d\varphi^{2}, 
\eq
where $b(\xi)$ is an arbitrary function of $\xi$ with $\xi \equiv T - R$.
Corresponding to the metric (\ref{eq2.2}), the energy-momentum tensor is given
by
\bq
\lb{eq2.3}
T^{+}_{\mu\nu} = \frac{b'(\xi)}{R} k_{\mu}k_{\nu},
\eq
where $k_{\mu}$ is a null vector, defined as
\bq
\lb{eq2.4}
k_{\mu} = \frac{1}{\sqrt{2}}\left(\delta^{T}_{\mu} - \delta^{R}_{\mu}\right),
\eq
which is the generator of the out-going radial null geodesic congruence
\cite{LW1994}. The presence of the out-going gravitational waves is indicated
by the only non-vanishing component of the Weyl tensor,
$C_{\mu\nu\lambda\sigma}$, given by \cite{WS1996}, 
\bq
\lb{eq2.3a}
\Psi_{0} \equiv
-C_{\mu\nu\lambda\sigma}L^{\mu}M^{\nu}L^{\lambda}M^{\sigma} = -
\frac{b'(\xi)}{2R}e^{b(\xi)},  
\eq
where $L^{\mu}$ and $M^{\mu}$ are null vectors, the definitions of which are
given by Eq.(8) in \cite{WS1996}.

 From Eqs.(\ref{eq2.1}) and (\ref{eq2.2}) we
find that the first junction conditions (\ref{eq6}) now reduce to
\bq
\lb{eq2.5}
d\tau = \left[1 - {r'}^{2}_{0}(t)\right]^{1/2} dt 
= e^{-b(\xi_{0})/2}\left[1 - {R'}^{2}_{0}(T)\right]^{1/2} dT,\;\;\;
r_{0}(t) = R_{0}(T),
\eq
where  $\xi_{0}$ is defined as $\xi_{0} = T - R_{0}(T)$. From the above
expressions we find 
\bq
\lb{eq2.6}
\left(\frac{dT}{dt}\right)^{2} = \frac{1}{\Delta}\equiv \left[{R'}^{2}_{0}(T) 
+ e^{-b(\xi_{0})}\left( 1 - {R'}^{2}_{0}(T)\right)\right]^{- 1},
\eq
which results in 
\bqn
\lb{eq2.7}
\frac{d^{2}T}{dt^{2}} &=& - \frac{1}{2\Delta^{2}}
\left\{2{R'}_{0}{R''}_{0} \right.\nb\\
& & \left. - e^{-b(\xi_{0})}
\left[b'(\xi_{0})(1 - {R'}_{0})(1 - {R'}^{2}_{0}) +
2{R'}_{0}{R''}_{0}\right] \right\}, \nb\\
{r''}_{0}(t) & =&
\frac{d^{2}T}{dt^{2}}{R'}_{0} + \left(\frac{dT}{dt}\right)^{2}{R''}_{0}\nb\\
&=& \frac{e^{-b(\xi_{0})}}{2\Delta^{2}}\left\{2{R''}_{0} +
b'(\xi_{0}){R'}_{0}(1 - {R'}_{0})(1 - {R'}^{2}_{0})\right\}. 
\eqn
Inserting Eqs.(\ref{eq9}) and the corresponding expressions for
$K^{-}_{ab}$ into Eq.(\ref{eq14}), and considering Eq.(\ref{eq2.7}), we find
\bqn 
\lb{eq2.8}
\rho &=& \frac{e^{b(\xi_{0})/2}}{\kappa {R}_{0}(1 - {R'}_{0}^{2})^{1/2}}
(\Delta - 1),\nb\\
p_{z} &=& \frac{e^{b(\xi_{0})/2}}{\kappa \Delta {R'}_{0}(1 -
{R'}_{0}^{2})^{3/2}} \left\{\Delta (1- \Delta )(1 - {R'}_{0}^{2}) - (1- \Delta
){R'}_{0}{R''}_{0}  \right.\nb\\
& & - \left. \frac{1}{2}b'(\xi_{0}){R'}_{0}({R'}_{0} - \Delta)(1 - {R'}_{0})(1 -
{R'}^{2}_{0}) \right\},\nb\\
p_{\varphi} &=& \frac{e^{b(\xi_{0})/2}}{\kappa \Delta (1 -
{R'}_{0}^{2})^{3/2}} \left\{(\Delta - 1){R''}_{0} \right.\nb\\
& & - \left. \frac{1}{2}b'(\xi_{0})({R'}_{0} - \Delta)(1 - {R'}_{0})(1 -
{R'}^{2}_{0}) \right\}.
\eqn

When the cylindrical thin shell is made of counter-rotating
dust particles, where half of the dust particles orbit around the symmetry
axis in a right-handed direction with angular momentum per unit rest mass $p$,
and the other half orbit in the opposite, left-handed direction with angular
momentum  per unit rest mass $ - p$, the surface energy-momentum tensor is
given by Eq.(\ref{eq13}) with $p_{z} = 0$ \cite{AT1992}. Thus, setting $p_{z}
= 0$ in Eq.(\ref{eq2.8}), we find
\bq
\lb{eq2.9}
{R''}_{0} = \frac{1 - {R'}_{0}^{2}}{{R'}_{0}}
\left\{\Delta  + \frac{1}{2}b'(\xi_{0}){R'}_{0}(1 - {R'}_{0})\frac{{R'}_{0} -
\Delta}{\Delta - 1}\right\},\; (p_{z} = 0).
\eq
Substituting the above expression into Eq.(\ref{eq2.8}), we obtain
\bq
\lb{eq2.10}
\rho = p_{\varphi} =  \frac{e^{b(\xi_{0})/2}}{\kappa {R}_{0}(1 -
{R'}_{0}^{2})^{1/2}} (\Delta - 1),\; (p_{z} = 0).
\eq


To further study the dynamics of the thin shell with $p_{z} = 0$, we need to
solve Eq.(\ref{eq2.9}), which is found too difficult to be done in the general
case. Therefore, in the following we shall consider a particular case in which
\bqn
\lb{eq2.13}
{R''}_{0} &=& \frac{1 - {R'}_{0}^{2}}{{R'}_{0}} \beta,\\
\lb{eq2.14}
\beta & = & \Delta  + \frac{1}{2}b'(\xi_{0}){R'}_{0}(1 - {R'}_{0})\frac{{R'}_{0} -
\Delta}{\Delta - 1}, 
\eqn
where $\beta$ is an arbitrary constant. Once ${R'}_{0}(T)$ is known,   the
function $b'(\xi)$ can be obtained from Eq.(\ref{eq2.14}) by quadrature.
Since we are mainly interested in the dynamics of the shell, in the
following we shall concentrate ourselves on  Eq.(\ref{eq2.13}). Integrating it
we find that  
\bq \lb{eq2.15} 
{R'}_{0}(T) = \pm \left(1 - e^{-2\beta T}\right)^{1/2},
\eq 
where the ``+" sign corresponds to an expanding shell, while the ``$-$" sign
corresponds to a contracting shell. In the following let us consider
the two cases separately.

\subsection{Expanding Thin Shells}

When the ``+" sign in Eq.(\ref{eq2.15}) is chosen, the integration of
it yields,  
\bq
\lb{eq2.16}
{R}_{0}(T) = (T + R_{min}) - \frac{1}{\beta}\left\{
 \left(1 - e^{-2\beta T}\right)^{1/2} - \ln\left[1 + \left(1 - e^{-2\beta
T}\right)^{1/2}\right]\right\},
\eq
where $R_{min}$ is an integration constant. When $\beta > 0$, we find that
\bqn
\lb{eq2.16a}
{R}_{0}(T) &=&
\cases{ 
R_{min}, & $T = 0$,\cr
+ \infty, & $T \rightarrow + \infty$,\cr}\nb\\
{R'}_{0}(T) &=&
\cases{
0, & $T = 0$,\cr
1, & $T \rightarrow + \infty$,
\cr}\; (\beta > 0),
\eqn
which shows that in this case the corresponding solution represents the
expansion of a thin shell made of counter-rotating dust particles. The
expansion starts from the moment $T = 0$ with the radius of the shell given by
$R(0) = R_{min}$. At this moment the shell has zero radial velocity but
infinitely large acceleration pointing outwards, as one can see from
Eq.(\ref{eq2.13}). Thus, the shell will expand until $T = + \infty$, where it
reaches to its maximal radius $R_{0}(+\infty) = + \infty$, with its radial
velocity ${R'}_{0}(+\infty) = +1$ and acceleration ${R''}_{0}(+\infty)
= 0$.

When $\beta < 0$, we find
\bqn
\lb{eq2.16b}
{R}_{0}(T) &=&
\cases{
R_{min}, & $T = 0$,\cr
0, & $T = - |T_{1}|$,\cr}
\nb\\
{R'}_{0}(T) &=&
\cases{
0, & $T = 0$,\cr
{\mbox{finite}}, & $T = - |T_{1}|$,
\cr} \; (\beta < 0).
\eqn
Thus, now the solution represents the expansion of a thin shell, starting from
zero radius at the moment $T = -|T_{1}|$. It expands until the moment $T =
0$, where its radial velocity and acceleration are given, respectively, by
${R'}_{0}(T = 0) = 0$  and ${R''}_{0}(T = 0) = -\infty$. Because of its huge
 acceleration that now points inwards, the shell will collapse
from this moment on, by following a process similar to that to be described
below.

\subsection{Collapsing Thin Shells}

When the ``$-$" sign in Eq.(\ref{eq2.15}) is chosen, 
we find that 
\bq
\lb{eq2.17}
{R}_{0}(T) = (R_{min} - T) + \frac{1}{\beta}\left\{
 \left(1 - e^{-2\beta T}\right)^{1/2} - \ln\left[1 + \left(1 - e^{-2\beta
T}\right)^{1/2}\right]\right\},
\eq
where $R_{min}$ is another integration constant. Thus, 
when $\beta > 0$, from Eq.(\ref{eq2.17}) we find that
\bqn
\lb{eq2.18}
{R}_{0}(T) &=&
\cases{
R_{min}, & $T = 0$,\cr
0, & $T = |T_{1}|$,
\cr}\nb\\
{R'}_{0}(T) &=&
\cases{
0, & $T = 0$,\cr
- \left(1 - e^{-2\beta |T_{1}|}\right)^{2}, & $T = |T_{1}|$,
\cr} \; (\beta > 0),
\eqn
which shows that now the shell starts to collapse at the moment $T =
0$ with zero radial velocity. The
collapse ends up at the moment $T = |T_{1}|$, where the whole shell contracts
into a line-like spacetime singularity, as Eqs.(\ref{eq2.3}) and
(\ref{eq2.10}) show. Therefore, unlike the case studied by AT \cite{AT1992},
in the present case the centrifugal forces of the courter-rotating dust
particles are not strong enough to prevent the collapse from forming a
spacetime singularity. 

When $\beta < 0$, from Eqs.(\ref{eq2.15}) and (\ref{eq2.17}), we find
\bqn
\lb{eq2.19}
{R}_{0}(T) &=&
\cases{
+ \infty \, & $T \rightarrow - \infty$,\cr
R_{min}, & $T = 0$,
\cr}\nb\\
{R'}_{0}(T) &=&
\cases{
-1 , & $T \rightarrow - \infty$,\cr
0, & $T = 0$,
\cr} \; (\beta < 0).
\eqn
Thus, in the present case the shell starts to
collapse at the moment $T = - \infty$ with its radius $R_{0}(- \infty) = +
\infty$ and its radial velocity ${R'}_{0}(- \infty) = -1$. As it collapses,
it is radiating massless particles and gravitational waves, as one can see
from Eqs.(\ref{eq2.3}) and (\ref{eq2.3a}). at the moment $T = 0$, it collapses
to its minimal radius $R_{0}(0) = R_{min}$, where its velocity becomes
zero. As far as $R_{min} \not= 0$, in this case no spacetime singularity is
formed, and the centrifugal forces of the dust particles now are strong enough
to halt the collapse. On the other hand, from Eq.(\ref{eq2.13}) we
can see that at T = 0 the acceleration of the shell becomes infinitely large
and points outwards. So, from this moment on, the shell will expand,  
by  following a process similar to that described in the last subsection. When
$R_{min} = 0$,  the centrifugal forces are still not strong
enough to prevent the shell from collapsing into a zero radius, whereby a
spacetime singularity is formed.

\section{Conclusions}

In this paper, the general formulas of a non-rotating dynamic thin shell that
connects two arbitrary cylindrical regions have been given in terms of the
metric coefficients and their first derivatives, using Israel's method. As an
application of these formulas, the dynamics of a thin shell made of
counter-rotating non-interacting particles, which emits both gravitational
waves and massless particles,  has been studied. It has been found that in
some cases the models represent  an expanding shell and others a collapsing
shell. For the collapsing shell, two possible final
states exist. In one case, after the shell collapses to a minimal non-zero
radius, it starts to expands, that is, the angular momentum of the dust
particles is strong enough to halt the collapse,  so that  a spacetime
singularity is prevented from forming on the symmetry axis.   However, in the
other case, the rotation is not strong enough to halt the collapse at a
finite non-zero radius, and as a result a spacetime singularity is formed
finally. These results are different from the ones
obtained by AT in the radiation-free case \cite{AT1992}, but similar to the
ones obtained by ST for rotating spheroids with radiation \cite{ST1992}.

\section*{Acknowledgment}

The financial assistance from CNPq and FAPERJ (AW)  is gratefully
acknowledged.

\end{document}